\documentclass[aps,prl,reprint,superscriptaddress,amsmath,amssymb,floatfix]{revtex4-2}
\usepackage{graphicx}
\usepackage{dcolumn}
\usepackage{bm}
\usepackage{xcolor}
\usepackage{chemformula}

\begin{document}
\title{Intrinsic antiferromagnetic half-metal and topological phases from the ferrovalley states of the sliding bilayer altermagnets}
\author{Shihao Zhang}
\email{zhangshh@hnu.edu.cn}
\affiliation{School of Physics and Electronics, Hunan University, Changsha 410082, China}
\begin{abstract}
Altermagnetism is characterized by non-relativistic spin splitting and zero total magnetic moments. In this work, intrinsic antiferromagnetic half-metallic and topological phases were discovered within the ferrovalley states of sliding bilayer altermagnets. 
Taking the V$_2$OSSe system for example, first-principles calculations indicate that the spin-dependent interlayer hopping in the ferrovalley state ensures a direct gap in one valley (one spin channel) and band inversion in the other valley (opposite spin channel), which is manifested as an intrinsic antiferromagnetic half-metal. The microscopic model and effective Hamiltonian employed in this research confirm the universal spin-dependent interlayer hopping in the sliding altermagnet bilayer.
Further calculations imply the existence of Chern insulator and gapless surface states in the sliding altermagnet bilayer. Adjusting the sliding direction can achieve the transition between different half-metals with conducting electrons of different spins, accompanied by the switching of gapless surface states of opposite spins. This work proposes a roadmap to achieve the intrinsic antiferromagnetic half-metals, laying the foundation for the potential applications of intrinsic antiferromagnetic half-metals and topological phases in spintronics.

\end{abstract}
\maketitle
\section{Introduction}
Antiferromagnetic half-metal is a unique class of materials that exhibits fully spin-polarized metallic behavior in one spin channel while maintaining an insulating gap in the opposite spin channel, coupled with zero net macroscopic magnetization due to exact compensation of local magnetic moments within the unit cell\cite{hu2012half,van1995half}. It combines the benefits of half-metallicity (100\% spin polarization at the Fermi level) and antiferromagnetism (no stray magnetic fields). This makes antiferromagnetic half-metal ideal for spintronic applications such as spin-polarized scanning tunneling microscopy tips\cite{van1995half}, domain-wall-free anchor layers in spin valves, and magnetic interference\cite{hu2012half,van1995half}. Some candidates have been proposed through theoretical studies, including perovskite cuprates Sr$_7$RbCaRe$_3$Cu$_4$O$_{24}$\cite{nie2008possible}, iron-pnictide derivatives BaCrFeAs$_2$\cite{hu2010half}, doped semiconductors\cite{akai2006half}, and Heusler alloys\cite{LUO20081797}. These systems leverage strong electron correlations, band engineering, and specific crystalline symmetries to achieve the required electronic and magnetic compensation. However, challenges encompass structural instability stemming from intricate compositions and doping, susceptibility to spin-orbit coupling, temperature-dependent magnetization compensation, and challenges in characterizing bulk properties without surface-sensitive experimental artifacts\cite{hu2012half}. These factors impede the practical application of antiferromagnetic half-metals. 

Recently, altermagnetism has emerged as a significant and distinct magnetic phase in condensed matter physics\cite{gRuO2,aCe4Sb3,bCe4Sb3,xiao2023spin,jiang2023enumeration,chen2023spin,ren2023enumeration,gao2023ai,qu2024extremely,tan2024bipolarized,guo2024valley,he2023nonrelativistic,okugawa2018weakly,hayami2019momentum,PhysRevB.101.220403,PhysRevB.102.144441,PhysRevB.75.115103}, bridging the characteristics of ferromagnets and antiferromagnets. Unlike ferromagnets, altermagnets exhibit zero net magnetization due to compensated spin sublattices, akin to antiferromagnets. However, their defining feature lies in the breaking of spin degeneracy without relying on spin-orbit coupling (SOC), resulting from crystallographic symmetries that prevent opposite spin sublattices from being interconverted by translation or inversion operations. This leads to a momentum-dependent spin-splitting of electronic bands. Such spin-splitting enables many unconventional phenomena, such as the crystal Hall effect\cite{aHalleffect,bHalleffect,cHalleffect,dHalleffect,eHalleffect,fHalleffect,gHalleffect,samanta2020crystal}, non-relativistic spin currents\cite{aSpincurrents,cSpincurrents,bSpincurrents,dSpincurrents,eSpincurrents,hHalleffect,naka2021perovskite}, and spin-splitting torques\cite{atorque,btorque}. Many material systems \cite{aRuO2,bRuO2,cRuO2,dRuO2,eRuO2,fRuO2,gRuO2,aMnF2,bMnF2,cMnF2,FeSb2,CrSb,CrSbexpriment,bCrSb,PhysRevLett.133.206401,liao2025direct,akCl,bkCl,Cr2SO,aRuF4,bRuF4,okugawa2018weakly,aV2Se2O,bV2Se2O,c2fq-hkk4,zhang2025sliding,li2025altermagnetism,peng2025ferroelastic} have been theoretically or experimentally identified as altermagnets, with their electronic structures revealing robust spin-polarized bands and lifted Kramers degeneracy. \textcolor{black}{Among the altermagnetic materials, numerous two-dimensional altermagnets have been reported in theoretical studies, such as V$_2$Se$_2$O, Fe$_2$MoS$_4$, Cr$_2$SO and CrO these Lieb-lattice monolayers\cite{xu2025chemical,aV2Se2O,bV2Se2O,li2025ferrovalley,Cr2SO,zhang2025sliding,v2te2o,Zou_2025,V2SeTeO,Fe2Se2O,fu2025strain}. Bilayer systems are also capable of exhibiting altermagnetism. Previous theoretical works about general stacking theory have formulated general methods for generating 2D altermagnetism via bilayer stacking configurations\cite{PhysRevLett.133.166701,PhysRevB.110.014442}. These developments position 2D altermagnets as promising candidates for spintronic applications, offering advantages of prolonged spin relaxation time and efficient spin manipulation through their unique combination of symmetry-protected properties and electric field tunability.}

In this work, we focused on the sliding bilayer altermagnet and utilized the ferrovalley states to attain the intrinsic antiferromagnetic half-metal and topological phases. In the following discussions, we studied the ferrovalley states in the sliding altermagnet Lieb-lattice bilayers, taking V$_2$OSSe bilayer as an example. Our first-principles calculations reveal that in the $x$-sliding V$_2$OSSe bilayer, spin-up energy bands have tiny direct bandgap. However, spin-down energy bands have the band inversion in the $Y$ valley. Similarly, spin-down energy bands have band inversion at the $X$ valley in the $y$-sliding V$_2$OSSe bilayer. We use the atomic model and effective model to underline the universal and important role of ferrovalley and spin-dependent interlayer hopping in the intrinsic antiferromagnetic half-metal. The band inversion indicates of non-trivial topological states in the sliding altermagnet bilayer. Further calculations about Berry curvatures and surface states verify the existence of a Chern insulator with gapless surface states in the sliding altermagnet bilayer. The adjustment of the sliding direction can realize the transition between different half-metals with conducting electrons of different spins, accompanied by the switching of gapless surface states of opposite spins. 

\begin{figure}
\includegraphics[width=1.0\linewidth]{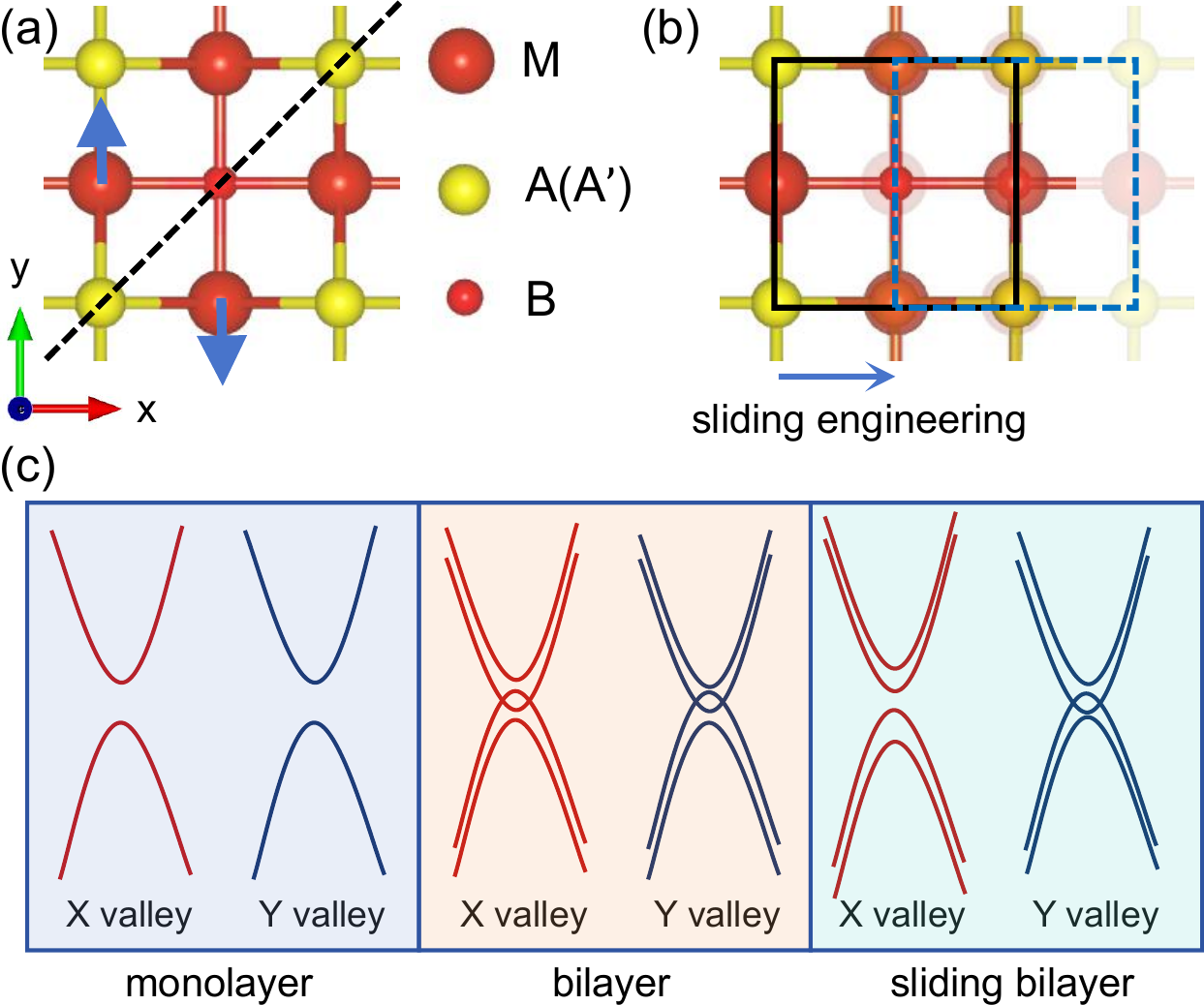}
\caption{(a) The structure of M$_2$A$_2$B or M$_2$AA$^{\prime}$B monolayer. Two magnetic sublattices carries opposite spin moments. The monolayer obeys the mirror symmetry, which is along dashed line direction. (b) The sliding engineering about altermagnet bilayer. The unit cells of two layers are marked by black solid lines and blue dashed lines, respectively. (c) The illustration about electronic structures of monolayer, AA-stacking bilayer and sliding bilayer altermagnets. The energy bands in different spin channels are marked by red and blue lines, respectively.}
\end{figure}

\section{Calculation methods}
We performed first-principles calculations by the Vienna Ab-initio Simulation Package (VASP)\cite{VASP} based on density functional theory. In all our calculations, we employed the Perdew-Becke-Ernzerhof functional within the generalized gradient approximation\cite{GGA} to describe the exchange-correlation potential. The plane wave cutoff energy was set to 600 eV. We used the optB88-vdW functional to describe the interlayer interaction. 

\section{Structure and altermagnetism} 
\begin{figure}
\includegraphics[width=1.0\linewidth]{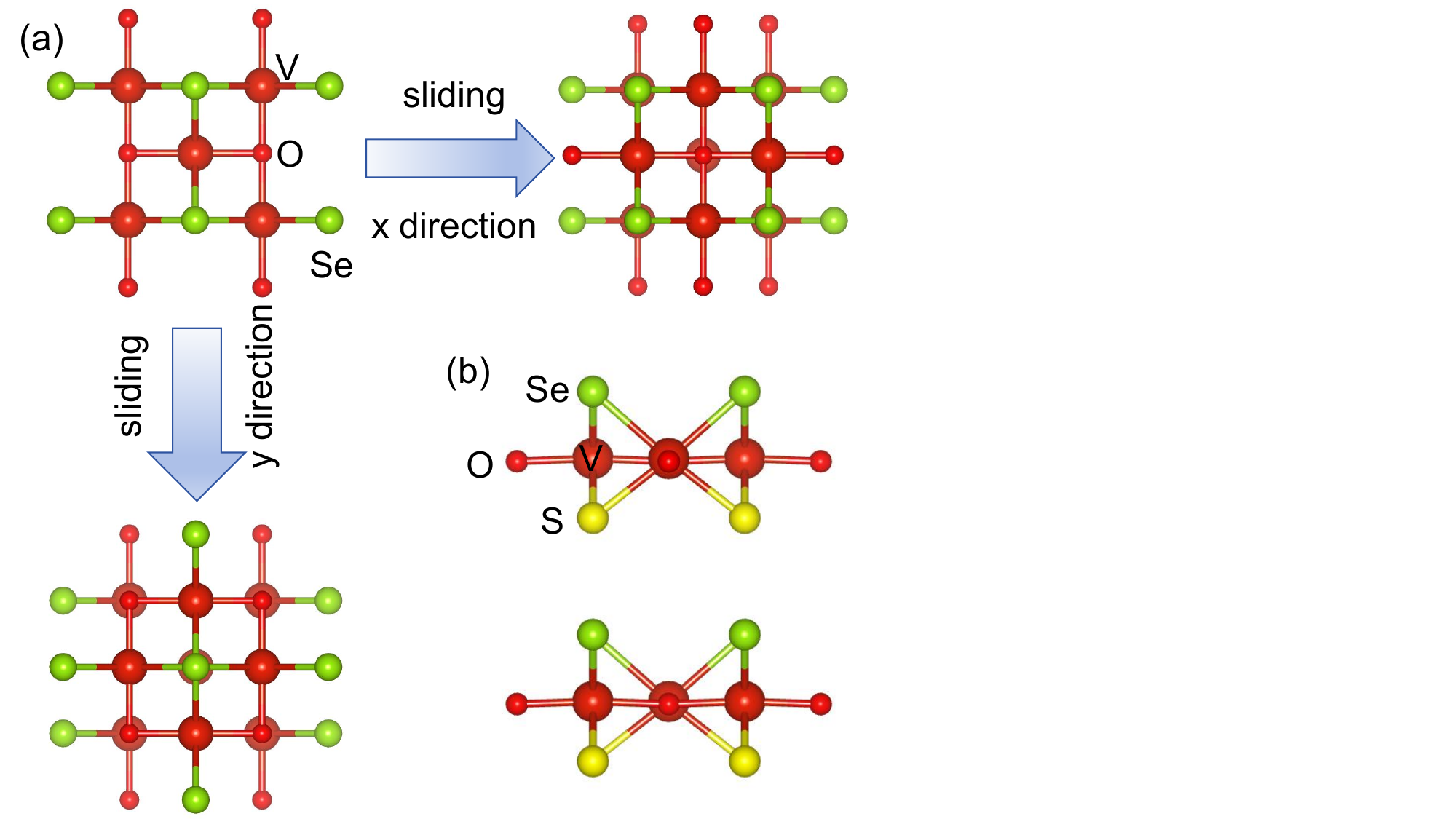}
\caption{(a) The sliding engineering about V$_2$OSSe bilayer. (b) The side view about sliding V$_2$OSSe bilayer.}
\end{figure}

Theoretical works have reported that many M$_2$A$_2$B and M$_2$AA$^{\prime}$B monolayers exhibit the altermagnetism\cite{xu2025chemical,aV2Se2O,bV2Se2O}. The M$_2$A$_2$B and M$_2$AA$^{\prime}$B monolayers possess a similar Lieb-lattice structure when viewed from the top. Here M is transition metal element, and A (A$^{\prime}$) or B denotes a group-IV element. In the monolayer, two M atoms and one B atom are situated within the same plane, and A atom or A$^{\prime}$ atom is positioned below and above the M-B atomic plane. In these monolayers, their valence band maximum and conduction band minimum are located at the X$(\pi/a,0)$ and Y$(0,\pi/a)$ points, where $a$ is in-plane lattice parameter of monolayer. These monolayers obey the $C_{4z}\mathcal{T}$ symmetry and mirror symmetry as depicted in the Fig.\,1, which results in non-relativistic spin-valley locking $E_{Y,\uparrow}=E_{X,\downarrow}$. The energy bands at the conduction band minimum in the X and $Y$ valley are mainly contributed by $d_{yz}$ and $d_{xz}$ orbitals of different sublattices, respectively. At the valence band maximum, the energy bands in both valleys stem from hybridized $d_{xy}$ and $p$ orbitals.

Now we focus on the AA-stacking M$_2$A$_2$B and M$_2$AA$^{\prime}$B bilayer. Owing to interlayer coupling, the bandgaps of the bilayers exhibit a significant reduction and may even transition to a metallic state, as depicted in the Fig.\,1(c). However, when one layer is fixed, and the other layer is sliding about half of lattice period along $x$ or $y$ direction, the mirror symmetry is destroyed and spin-valley locking vanishes in the sliding bilayer. Thus, sliding bilayer system show the \textcolor{black}{ferrovalley or valley-imbalance} state as shown in the Fig.\,1\cite{li2025ferrovalley}, in which different spin channels possess distinct bandgap. When one layer is sliding along $x$ direction, the spin-down electrons have smaller bandgap in the $Y$ valley. Conversely, if sliding direction is switched to $y$ direction, spin-up electrons feature a smaller bandgap in the $X$ valley. 

\begin{figure}
\includegraphics[width=1.0\linewidth]{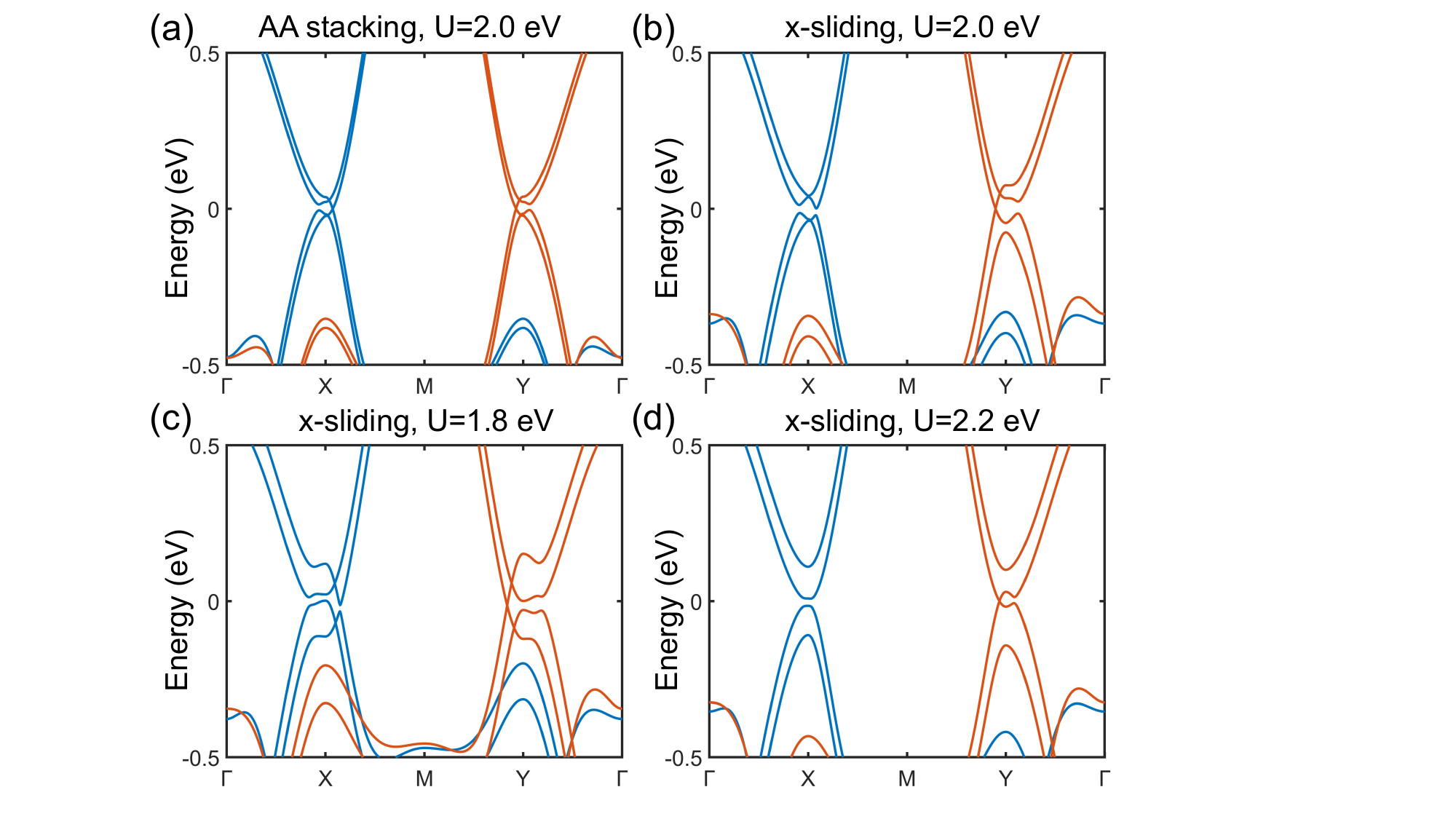}
\caption{The first-principles energy bands of (a) AA-stacking and (b-d) x-sliding bilayers. The Hubbard parameter of V atom is set to 2.0\,eV in the (a,b), 1.8\,eV in the (c) and 2.2\,eV in the (d). The energy bands of different spin are marked by red and blue lines, respectively.}
\end{figure}

\section{Non-relativistic ferrovalley electronic structures} 
Because the V$_2$OSSe monolayer has small bandgap of only 0.17\,eV, we focus on the electronic structure of V$_2$OSSe system, where band inversion is easy to be realized. In this monolayer, S and Se atoms are situated at the A sites. \textcolor{black}{The structure of V$_2$OSSe bilayer are present in the Fig.\,2.} In the Fig.\,3, we present the electronic structures of V$_2$OSSe bilayer. Our calculations as shown in the Fig.\,S1 and Table.\,S1 of supplementary materials reveal that sliding bilayers has lower total energies than AA-stacking bilayers \textcolor{black}{by 78.1\,meV}, and each layer still holds altermagnetism. \textcolor{black}{During the sliding engineering, the barrier reaches 278.6\,meV as shown in the Fig.\,S2. No experiments have been conducted on the V$_2$OSSe system; however, angle-resolved photoemission spectroscopy (ARPES) experiments have been carried out on the similar system, the K$_2$V$_2$Se$_2$O bulk\cite{Jiang2025}. The energy bands of the K$_2$V$_2$Se$_2$O bulk were calculated using the HSE06 functional\cite{HSE06} (which typically provides higher-level theoretical data) and compared with the experimental results. The results are presented in Fig. S3. Through this comparison, it can be observed that the HSE06 functional overestimates the energy band gaps and fails to capture the Fermi pocket near the X (Y) point. Therefore, the K$_2$V$_2$Se$_2$O system exhibits a weak correlation effect. Based on this comparison, it is hypothesized that the V$_2$OSSe system, which is similar to K$_2$V$_2$Se$_2$O, also demonstrates a weak correlation effect.} In the AA-stacking bilayer, $C_{4z}\mathcal{T}$ symmetry ensures that the energy bands of different spins undergo band inversion in both the $X$ and $Y$ valleys. In the x-sliding bilayer, band inversion persists in both valleys. However, due to broken mirror symmetry in the sliding bilayer, a direct bandgap emerges in the $X$ valley, and electrons with opposite spins in the $Y$ valley keep conducting, as illustrated in the Fig.\,3(b). When the bandgap is further reduced by decreasing the Hubbard parameter of M atoms, the direct bandgap in the $X$ valley remains, yet the global bandgap vanishes due to the strong hybridization between valence bands and conduction bands. Conversely, when the bandgap is increased, the band inversion in the $X$ valley disappears, while the band inversion in the $Y$ valley is still maintained. 

To unravel the physical origin of electronic structure in the sliding altermagnetic M$_2$A$_2$B or M$_2$AA$^{\prime}$B bilayer, the microscopic dicussions about detailed interlayer orbital hopping are necessary. Within Slater-Koster tight-binding model, the interlayer coupling between valence band's $d_{xy}$ orbital and conduction band's $d_{yz}/d_{xz}$ orbitals can be written as

\begin{equation}
\begin{pmatrix}
E_{xy,yz} \\
E_{xy,xz}
\end{pmatrix}
=
\begin{pmatrix}
3l m^{2} n & ln(1-4m^{2}) & ln(m^{2}-1) \\
3l^{2} m n & m n (1-l^{2}) & m n (l^{2}-1)
\end{pmatrix}
\begin{pmatrix}
V_{dd\sigma} \\
V_{dd\pi} \\
V_{dd\delta}
\end{pmatrix}
\end{equation}
Here the distance between two orbitals is $\bm{d}=(x,y,z)=|\bm{d}|(l,m,n)$, where $l$, $m$, and $n$ represent the directional cosines. In the AA-stacking bilayer, interlayer hopping direction satisfies $l=m=0$, indicated of $E_{xy,yz}=E_{xy,xz}=0$. Thus, the overlapping bands in both $X$ and $Y$ valleys are not affected by the interlayer hopping. However, in the $x$-sliding bilayer, interlayer hopping direction corresponds to $m = 0$. Under this condition, we obtain $E_{xy,yz}=ln(V_{dd\pi}-V_{dd\delta})$ and $E_{xy,xz}=0$. The non-zero value of $E_{xy,yz}$ creates a direct energy gap in the overlapped valence and conduction bands in the $X$ valley. Meanwhile, the zero value of $E_{xy,xz}$ suggests that the interlayer hopping has no impact on the band inversion in the $Y$ valley. Regarding the $y$-sliding bilayer, interlayer hopping with $l=0$ results in $E_{xy,xz}=mn(V_{dd\pi}-V_{dd\delta})$ and $E_{xy,yz}=0$, which induces a direct gap in the $Y$ valley. Thus, as shown in the Fig.\,3, interlayer orbital coupling selectively open the non-relativistic bandgap in the overlapped bands in one valley, and keep the other valley's electrons conducting. 
Since different orbitals of different magnetic sublattices contribute to the energy bands of different spins, the interlayer orbital coupling gives rise to spin-dependent interlayer coupling, as illustrated in the Fig.\,4. \textcolor{black}{Our Wannier-function analysis reveal that the strength of interlayer hopping between $d_{xy}$ and $d_{yz}$ in the spin-up channel reaches 10.2\,meV, but the hopping between $d_{xy}$ and $d_{xz}$ in the spin-down channel is only 3.5\,meV.} Especially, the spin-dependent interlayer coupling is universal in the M$_2$A$_2$B and M$_2$AA$^{\prime}$B altermagnet systems\cite{zhang2025sliding}, which is beneficial to build the intrinsic antiferromagnetic half-metal. 

\begin{figure}
\includegraphics[width=1.0\linewidth]{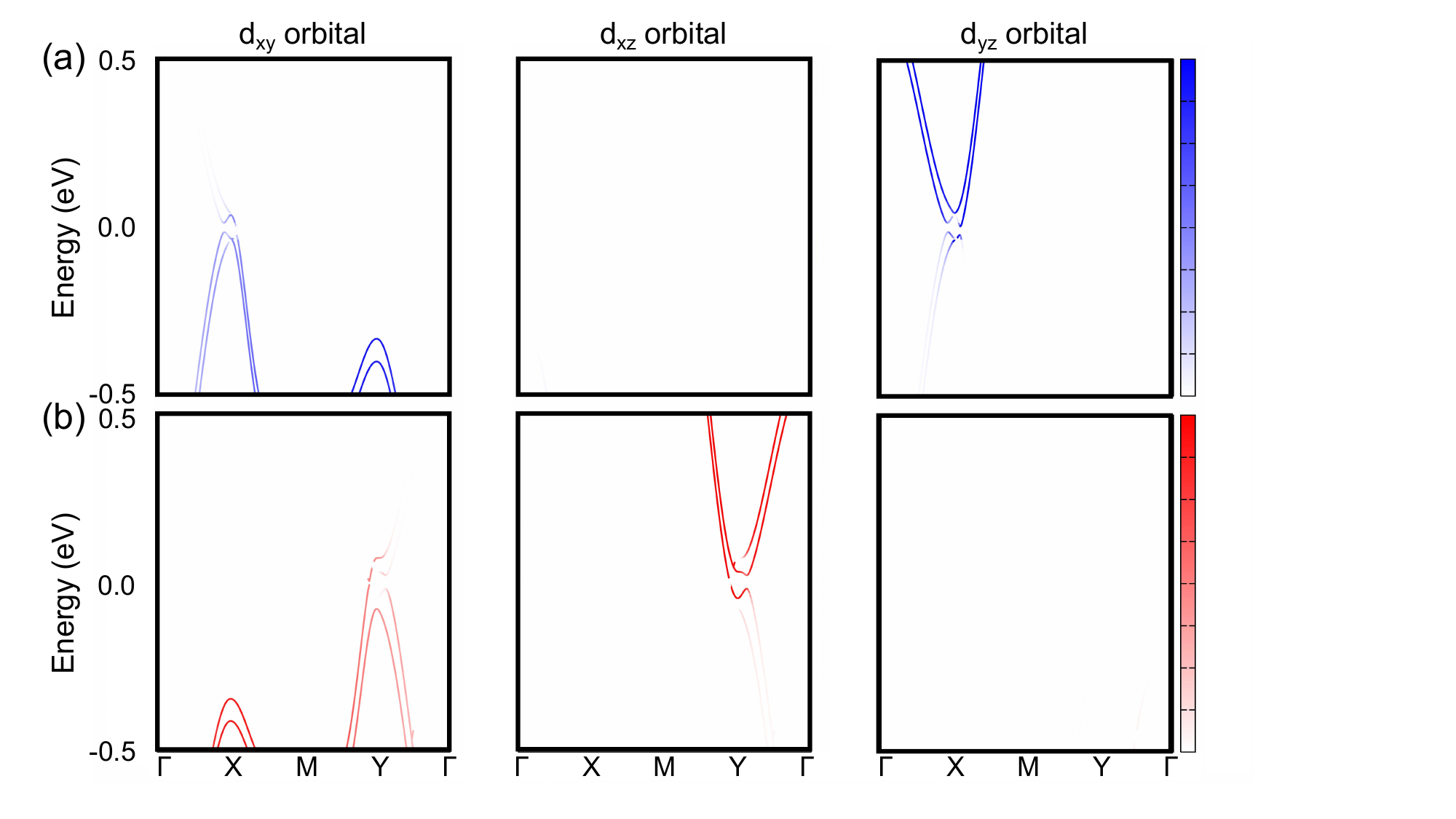}
\caption{The orbital-resolved energy bands of x-sliding bilayer V$_2$OSSe in the (a) spin-up channel and (b) spin-down channel. Here Hubbard parameter is set to 2\,eV.}
\end{figure}

Based on the discussions presented above, a simple effective model can be employed to characterize the ferrovalley states in the sliding M$_2$A$_2$B or M$_2$AA$^{\prime}$B bilayer. The discussions begin with the effective Hamiltonian of M$_2$A$_2$B or M$_2$AA$^{\prime}$B monolayer\cite{PhysRevB.110.144412,fu2025strain}
\begin{equation}
    \mathcal{H}_{0} = \left[ u\sigma_{z} + D (\cos k_{x} - \cos k_{y}) \right]\tau_{z} + t \cos \frac{k_{x}}{2} \cos \frac{k_{y}}{2} \tau_{x}
\end{equation}
Here $\tau$ and $\sigma$ are Pauli matrices defined in the sublattice and spin space. \textcolor{black}{$u$ denotes the antiferromagnetic staggered field, and $D$ describes the next-nearest-neighboring hopping term.} In the AA-stacking case, the Hamiltonian can be written as $\mathcal{H}_{AA} = \mathcal{H}_{0}l_0+t'l_x\tau_0\sigma_0$. Here $l$ denote Pauli matrices defined in the layer space, and $t'l_x\tau_0\sigma_0$ represents the interlayer hopping term. In the x-sliding bilayer, we consider the spin-dependent interlayer coupling and the Hamiltonian became
\begin{equation}
    \mathcal{H}_{x-sliding} = \mathcal{H}_{0}l_0+\cos k_x (t'l_x\tau_0\sigma_0+t''l_x\tau_0\sigma_z)
\end{equation}
\textcolor{black}{Here $t''l_x\tau_0\sigma_z$ term denotes aforementioned spin-dependent interlayer coupling, which originates from interlayer orbital coupling via orbital-spin locking. Here this interlayer orbital coupling is sensitive to the hopping orientation due to the anisotropic $d$ orbitals.} Then we can obtain that the direct gap at the two valleys,
\begin{equation}
\left\{ 
\begin{matrix} 
\Delta_{X} = 2|u - 2D| - 2\left| t^{\prime} - t^{\prime\prime} \right| \\ 
\Delta_{Y} = 2|u - 2D| - 2\left| t^{\prime} + t^{\prime\prime} \right| 
\end{matrix} 
\right.
\end{equation}
Thus, the spin-dependent interlayer coupling $t''$ gives rise to the ferrovalley state. \textcolor{black}{Because the conduction bands originate from hybridizated d$_{xy}$ and $p$ orbitals, the interlayer hopping between $p$ orbitals of non-magnetic atoms provides the non-zero spin-independent term of interlayer coupling.} If \textcolor{black}{spin-dependent term} $t^{\prime\prime}$ is zero, the gaps at two valleys are equal. In the x-sliding bilayer, the solutions can be classified into four possible phases,
\begin{equation}
\left\{ 
\begin{aligned} 
&\text{Phase-I}:   & 2|u-2D| &\le |t'-t''|\\
&\text{Phase-II}:  & |t'-t''| &< 2|u-2D| < 2|t'-t''|\\
&\text{Phase-III}: & 2|t'-t''| &< 2|u-2D| < 2|t'+t''|\\
&\text{Phase-IV}:  & 2|u-2D| &> 2|t'+t''|
\end{aligned} 
\right.
\end{equation}

\begin{figure}
\includegraphics[width=1.0\linewidth]{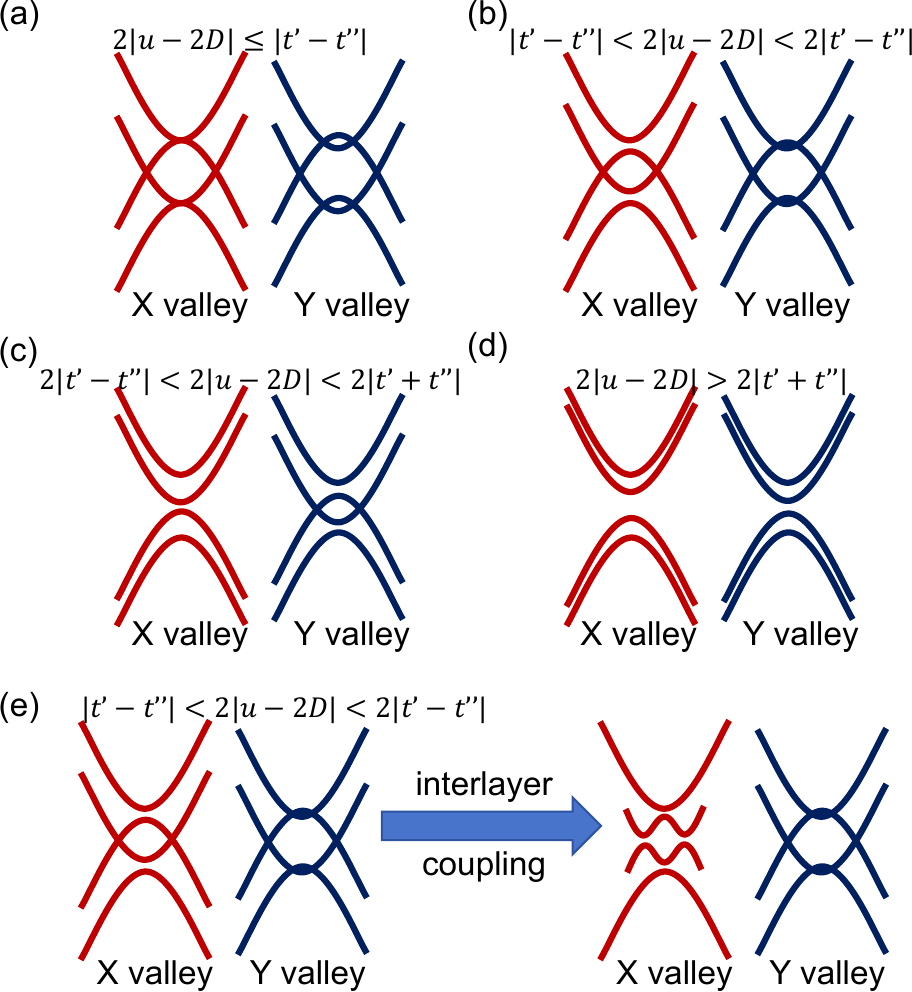}
\caption{Four possible phases (a-d) in the ferrovalley states in the x-sliding bilayer. The energy bands in different spin channels are marked by red and blue lines, respectively. Here interlayer orbital hopping is not taken into consideration. Only phase-III(c) is intrinsic antiferromagnetic half-metal. (e) While we consider detailed interlayer orbital hopping, only phase-II (b) and phase-III (c) are intrinsic antiferromagnetic half-metal.}
\end{figure}

\begin{figure*}
\includegraphics[width=1.0\linewidth]{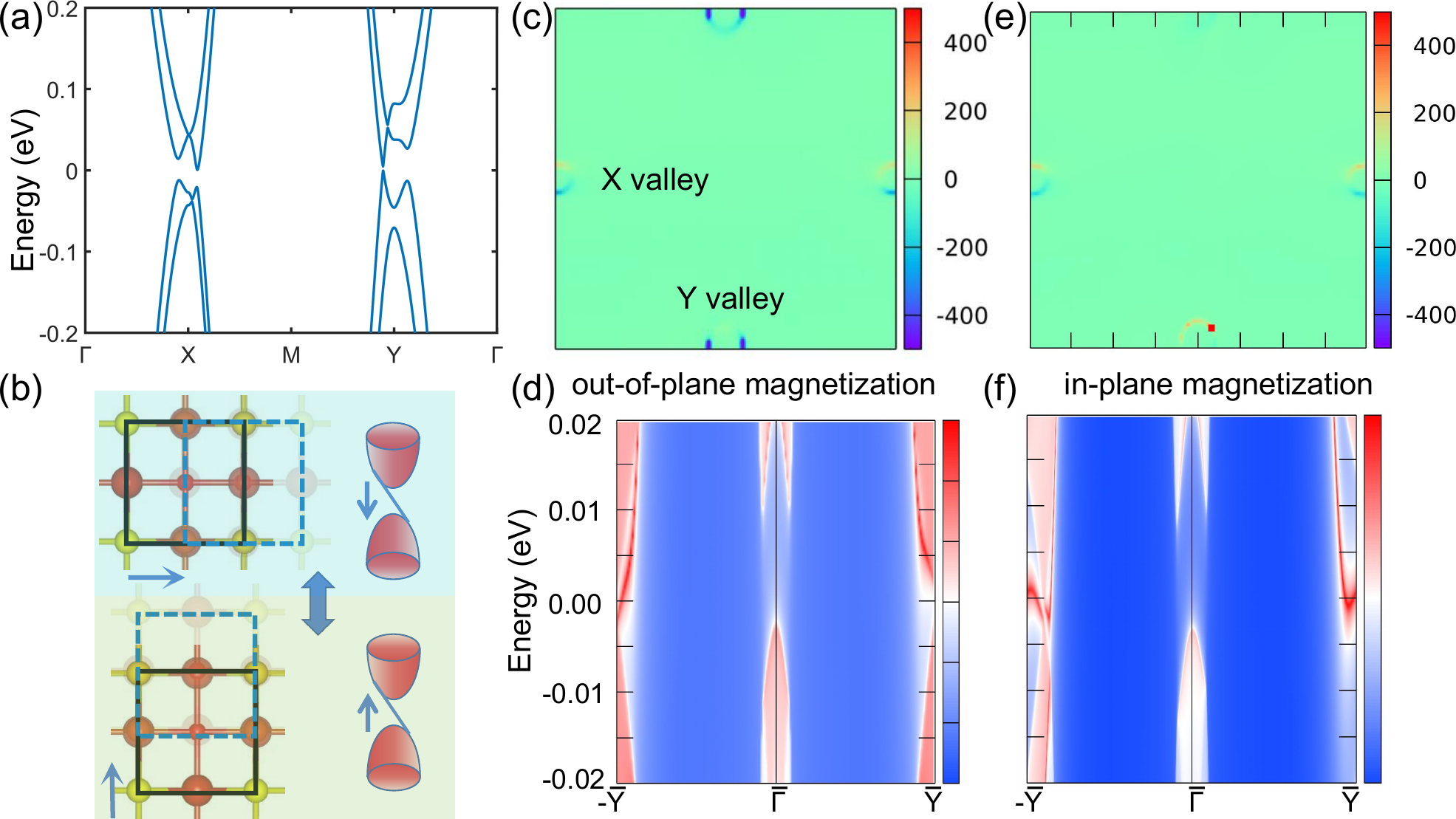}
\caption{(a) The energy bands with SOC effect in the x-sliding V$_2$OSSe bilayer with Hubbard parameter U=2.0\,eV. (b) The illustration about sliding engineering on the half-metal and gapless surface states. (c) The Berry curvatures in the Brillouin zone and (d) the gapless surface states in the electronic structures of (100) surface of the out-of-plane magnetization state. (e) The Berry curvatures in the Brillouin zone and (f) the gapless surface states of the in-plane magnetization phase. }
\end{figure*}

As depicted in Fig.\,5, phase-I possesses many crossing points and denotes a strong hybridization between valence bands and conduction bands within the $X$ valley. Consequently, it is difficult for interlayer orbital coupling to create a global band gap in the $X$ valley as shown in the Fig.\,3(c), leading to metallic phases in both spin channels. In phase-II, band inversions occur in both valleys. Nevertheless, spin-dependent interlayer hoppings can generate a global bandgap in the $X$ valley while exerting no influence on the band inversion in the $Y$ valley. Therefore, phase-II is characterized as an antiferromagnetic half-metal, referring to Fig.\,3(b). Regarding phase-III, the band inversion in the $X$ valley vanishes, yet it persists in the $Y$ valley, rendering phase-III an antiferromagnetic half-metal as shown in the Fig.\,3(d). In phase-IV, the band inversions in both valleys disappear, and the sliding altermagnet bilayer manifests as a ferrovalley semiconductor. In summary, ferrovalley states form the antiferromagnetic half-metal as phase-III, and interlayer orbital hoppings between $d_{xy}$ and $d_{yz}/d_{xz}$ orbitals expand the antiferromagnetic half-metal to include phase-II.

\textcolor{black}{We turn to the discussion about realistic material sliding bilayer V$_2$OSSe. When Hubbard parameter meet $U\leq$1.8\,eV, the sliding bilayer exhibit Phase-I electronic phase. While $1.8\,\rm{eV}< U\leq$2.4\,eV, slding bilayer V$_2$OSSe enters the Phase-II or Phase-III phase and thus behaves antiferromagnetic half-metal. When $U\geq $2.6\,eV, the sliding bilayer V$_2$OSSe shows the Phase-IV and becomes the antiferromagnetic ferrovalley semiconductor.}

\textcolor{black}{Although the effective Hamiltonian in Eq. (2) fails to capture the orbital character of the valence and conduction bands, we can utilize the relevant $t^\prime$ and $t^{\prime\prime}$ to define the average interlayer coupling and spin-dependent interlayer coupling. Here, we derive the strength of the interlayer hopping $t_1$ between $d_{xy}$ and $d_{yz}$ in the spin-up channel, as well as that of the inter-layer hopping $t_2$ between $d_{xy}$ and $d_{xz}$ in the spin-down channel. Then we define $t^\prime= (t_1 + t_2)/2$ and $t^{\prime\prime}= (t_1 - t_2)/2$. Thus, the strengths of $t^\prime$ and $t^{\prime\prime}$ in the sliding V$_2$OSSe bilayer attain 6.9 meV and 3.3 meV, respectively.}

\section{Nontrivial topological electronic structures with SOC effect}
We further performed first-principles calculations about $x$-sliding V$_2$OSSe bilayer with SOC effect. As shown in the Fig.\,6(a), the SOC effect open a bandgap of about 1\,meV at the overlapping bands. The calculated non-zero Berry curvatures are concentrated in two valleys as depicted in the Fig.\,6(c). However, the Berry curvatures in the $X$ valley provide zero total Chern number. In the $Y$ valley, the Berry curvatures contribute to $C=1$ nontrivial topological phase. Accompanied with gapless surface states shown in the Fig.\,6(d), the x-sliding altermagnet bilayer \textcolor{black}{with out-of-plane magnetization} exhibits as a Chern insulator. 

The x-sliding bilayer and y-sliding bilayer can be connected by $C_{4z}\mathcal{T}$ operation. Thus, the gapless surface state in the $Y$ valley and one spin channel of the x-sliding bilayer can be switched to that in the $X$ valley and the other spin channel of y-sliding bilayer under $C_{4z}\mathcal{T}$ operation. Thus, we can achieve the switch between different half-metals with conducting electrons in the opposite spin channel by tuning the sliding direction as illustrated in the Fig.\,6(b).

\begin{figure*}
\includegraphics[width=1.0\linewidth]{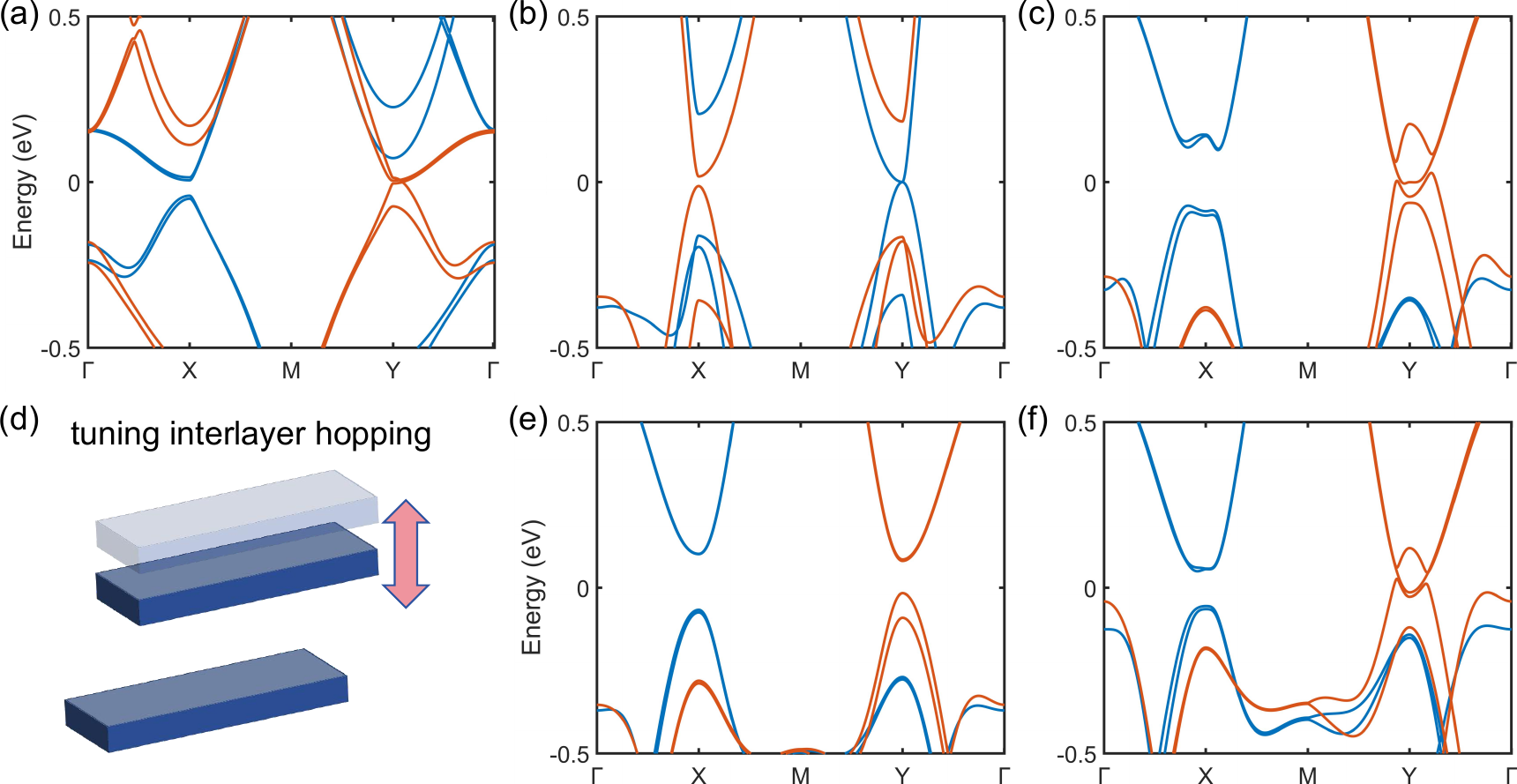}
\caption{The energy bands of the (a) sliding Fe$_2$MoS$_4$ bilayer with no Hubbard parameter, (b) sliding Cr$_2$SO bilayer with Hubbard parameter of 2.6\,eV, (c) sliding V$_2$Se$_2$O bilayer with Hubbard parameter of 2.0\,eV. (d) Thu illustration of antiferromagnetic half-metal realization by tuning the interlayer coupling of sliding altermagnetic bilayer. (e) The energy bands of the sliding V$_2$S$_2$O bilayer with Hubbard parameter of 2.0\,eV. (f) The energy bands of the sliding V$_2$S$_2$O bilayer when the interlayer distance is decreased by 0.7\AA{} under out-of-plane pressure. The energy bands of different spin are marked by red and blue lines, respectively.}
\end{figure*}

\textcolor{black}{Our first-principles calculations show that the total energy of in-plane magnetization in the sliding bilayer is lower than that of out-of-plane magnetization by 0.36$\sim$1.50 meV per magnetic atom. The spin orientation may be sensitive to charge transfer, orbital hybridization or strain effect from the substrate in the experiments. For example, while slding bilayer V$_2$OSSe is placed on the SrTiO$_3$ substrate and keeps no lattice strain, with the corresponding model presented in the Fig.\,S4 of the supplementary materials, our calculation shows that the magnetic anisotropy energy is significantly reduced to only 50\,$\mu$eV per magnetic atom. Moreover, the easy axis of magnetization transitions to the out-of-plane direction. As depicted in Fig.\,6(e,f) and Fig.\,S5, the sliding bilayer with in-plane magnetization remains a Chern insulator. Therefore, the topological characteristic of the sliding bilayer is robust with respect to the magnetization direction.  }

\textcolor{black}{The minimal SOC gap can be attributed to the feeble SOC effect of vanadium atoms, which is observable at liquid helium temperature. When the temperature is higher than that of liquid helium but lower than the Néel temperature, the SOC gap is overshadowed by thermal fluctuations. Consequently, the sliding bilayer exhibits the behavior of an antiferromagnetic half-metal without the influence of the SOC effect. We defer the exploration of materials with a large SOC gap in antiferromagnetic half-metals to future research.}

\section{Discussions} 

\textcolor{black}{We discuss the structural stability of V$_2$OSSe bilayer. Our calculations show that the total energy of V$_2$OSSe is higher than that of the phase-separated mixture of V$_2$OS$_2$ and V$_2$OSe$_2$ by 211.2 meV per formula unit, which makes possible spontaneous phase separation into V$_2$OS$_2$ and V$_2$OSe$_2$ thermodynamically favorable under equilibrium conditions. Thus, Janus structure design is adopted to kinetically and structurally stabilize the monophase V$_2$OSSe. The core strategy relies on constructing asymmetric out-of-plane atomic arrangement with ordered S and Se occupancy, rather than random atomic mixing. Two feasible synthetic routes are proposed: pre-synthesize pristine V$_2$OS$_2$ monolayer first, then selectively substitute partial surface S atoms with Se via low-temperature plasma selenization; alternatively, prepare V$_2$OSe$_2$ as the template and replace surface Se with S through mild sulfuration to form Janus V$_2$OSSe with asymmetric S/Se distribution on two sides. The intrinsic out-of-plane dipole moment and broken inversion symmetry of the Janus configuration introduce extra structural stabilization energy induced by substrate, which compensates the thermodynamic energy disadvantage of V$_2$OSSe. Meanwhile, the asymmetric atomic arrangement suppresses long-range diffusion and segregation of S and Se atoms, kinetically blocking the phase separation pathway toward V$_2$OS$_2$ and V$_2$OSe$_2$. Combined with low-temperature growth and rapid quenching, the Janus structural confinement further restricts lattice relaxation, enabling the stable synthesis of single-phase metastable V$_2$OSSe.}

We note that uniaxial strain can induce a ferrovalley state in the altermagnet monolayer by breaking the mirror symmetry\cite{fu2025strain}. However, altermagnetism is contingent upon the intralayer magnetic coupling among magnetic sublattices, which is sensitive to in-plane uniaxial stress. \textcolor{black}{As shown in the Table.\,S2, the uniaxial stress can lead the system to transition from antiferromagnetism into ferrimagnetism because the $C_{4z}\mathcal{T}$ symmetry between different magnetic sublattice is destroyed by uniaxial stress due to lattice deformation}. Given that the interlayer magnetic coupling is significantly smaller than the intralayer exchange coupling, sliding engineering is viable as it can preserve the intralayer altermagnetism. Previous theoretical works also have reported that external electric field can realize the ferrovalley state\cite{guo2024valley,PhysRevLett.133.056401}, which differs from our work. Our work proposed an intrinsic mechanism for ferrovalley electronic structures through sliding engineering. Without the requirement of an external field or uniaxial strain, this proposal facilitates the realization of the ferrovalley phase and even the antiferromagnetic half-metal. 

The sliding engineering provides feasible platform to achieve the antiferromagnetic half-metal and nontrivial topological phase. If sliding bilayer fails to open a global gap in one valley, the interlayer distance can be increased to diminish the interlayer coupling, thereby moderately reducing the negative bandgap in this valley.

It should be noted that the calculated bandgaps are always sensitive to Hubbard parameters and van der Waals functionals \textcolor{black}{as shown in the Fig.\,S6}. However, in this work, our proposal is not limited to V$_2$OSSe system. We propose that by utilizing an M$_2$A$_2$B or M$_2$AA$^{\prime}$B altermagnetic monolayer with a small bandgap, the intrinsic antiferromagnetic half-metal state can be attained through sliding engineering along with ferrovalley states. Moreover, nontrivial topological phases can be realized via band inversion in these sliding bilayers. 

\textcolor{black}{Antiferromagnetic half-metals are not only realized in the sliding M$_2$A$_2$B or M$_2$AA$^{\prime}$B bilayers, but also exist in other Lieb-like lattice altermagnetic sliding bilayers. As shown in Fig. 7(a-c), the sliding Fe$_2$MoS$_4$, Cr$_2$SO, and V$_2$Se$_2$O bilayers may all exhibit antiferromagnetic half-metallic properties. It should be noted that first-principles calculations often overestimate or underestimate the bandgap, which restricts the conclusions regarding the universality of antiferromagnetic half-metals. For instance, the sliding Cr$_2$SO bilayer might possess an antiferromagnetic metal rather than a half-metal under a Hubbard parameter of 2.0\,eV. To overcome the limitation associated with the uncertainty of the bandgap in first-principles calculations, a universal method for obtaining an antiferromagnetic half-metal is also sought. The calculations in Fig. 7(d-f) show that adjusting the interlayer distance between the sliding bilayers can also achieve an antiferromagnetic half-metal in the altermagnetic bilayer with a small bandgap. We also calculated the spin-dependent interlayer coupling in different sliding bilayers, as depicted in Fig.\,7. The results are presented in Table 1, which indicate that the spin-dependent interlayer coupling in different sliding bilayers can attain several meV. This adjustment method can expand the accessible region of the antiferromagnetic half-metal in the material phase.}

\begin{table}[htbp]
  \centering
  \caption{The calculated average interlayer coupling strength $|t_1|$ and spin-dependent interlayer coupling strength $|t_2|$ in different sliding bilayers. All coupling strengths are in unit of meV.}
  \begin{tabular*}{\linewidth}{@{\extracolsep{\fill}} lcc }
    \hline \hline
    Material         & $|t_1|$ & $|t_2|$ \\
    \hline
    V$_2$OSSe        & 6.9     & 3.3     \\
  Fe$_2$MoS$_4$    & 4.2    & 5.1    \\
  Cr$_2$SO         & 2.3    & 1.1     \\
  V$_2$Se$_2$O     & 1.7     & 4.4     \\
  V$_2$S$_2$O      & 3.6     & 2.2     \\
  V$_2$S$_2$O (compressed) & 0.5 & 3.5 \\
    \hline \hline
  \end{tabular*}
\end{table}

\section{Conclusions} 
In this study, we employed the altermagnet monolayer featuring a small bandgap to construct the bilayer system. The inter-layer hopping phenomenon reduces the bandgaps, and sliding engineering gives rise to ferrovalley states. Taking the V$_2$OSSe system as an example, our first-principles calculations demonstrate that the spin-dependent inter-layer hopping in the ferrovalley state guarantees a direct gap in one valley (one spin channel) and band inversion in the other valley (opposite spin channel), which manifests as an intrinsic antiferromagnetic half-metal. Our microscopic model and effective Hamiltonian verify the universal spin-dependent inter-layer hopping in the sliding altermagnet bilayer. Further calculations suggest the existence of Chern insulator and gapless surface states in the sliding altermagnet bilayer. The adjustment of the sliding direction can realize the switch between different half-metals with conducting electrons of different spins, accompanied by the switching of gapless surface states of opposite spins. Our research pave the way for the promising applications of intrinsic antiferromagnetic half-metals and topological phases in spintronics. 

\section{acknowledgments}
We thank Dr. X. Zhang and Dr. H.-Y. Ma for helpful discussions and comments. This work was supported by the National Natural Science Foundation of China (No. 12304217), the National Key Research and Development Program of China (No. 2024YFA1410300), the Natural Science Foundation of Hunan Province (No. 2025JJ60002) and the Fundamental Research Funds for the Central Universities from China (No. 531119200247).



\section{Author Contributions}
Shihao Zhang performed all calculations, prepared all figures and wrote the main manuscript text.

\section{Competing Interests}
The authors declare no competing interests.

\section{Data availability}
The data that support the findings of this study are available from the corresponding authors upon reasonable request.
%

\end{document}